\documentclass{article}

\usepackage{amsfonts}
\usepackage{epsfig}

\begin{document}

\title{\textsc{City Space Syntax as a Complex Network}}

\vspace{1cm}

\author{ D. Volchenkov and Ph. Blanchard
\vspace{0.5cm}\\
{\it  BiBoS, University Bielefeld, Postfach 100131,}\\
{\it D-33501, Bielefeld, Germany} \\
{\it Phone: +49 (0)521 / 106-2972 } \\
{\it Fax: +49 (0)521 / 106-6455 } \\
{\it E-Mail: VOLCHENK@Physik.Uni-Bielefeld.DE}}
\large

\date{\today}
\maketitle

\begin{abstract}
The encoding of cities into non-planar dual graphs
reveals their complex
structure.
 We investigate the statistics of
the typical space syntax measures
for the five different compact urban patterns.
Universal statistical behavior of space syntax measures
 uncovers the universality of
the city creation
mechanism.
\end{abstract}

\vspace{0.5cm}

\leftline{\textbf{ PACS codes: 89.75.Fb, 89.75.-k, 89.90.+n} }
 \vspace{0.5cm}

\leftline{\textbf{ Keywords: Complex networks, city space syntax} }

\section{Paradigm of a city}
\label{Sec:Introduction}
 \noindent

Although, nowadays the majority of people live in cities \cite{Crane}
there is no one standard international definition of a city: the term may
be used either for a town possessing city status; for an urban locality
exceeding an arbitrary population size; for a town dominating other towns
with particular regional economic or administrative significance. In most
 parts of the world, cities are generally substantial and nearly always
 have an urban core, but in the US many incorporated areas which have a
  very modest population, or a suburban or even mostly rural character,
   are designated as cities.

Cities have  often been compared with
biological entities \cite{Miller}. The implication that social organizations and dynamics
relating urbanization to economic development and knowledge
production
are extensions of biology, satisfying similar principles and
constraints has a strong empirical ground.
Almost all physiological characteristics of
biological organisms scale with the mass of their bodies, $M$,  as a power law.
For example, it has been reported in \cite{Enquist}
that the curve describing the power, $B$, required to sustain a living organism
is a straight line in the logarithmic scale,
\begin{equation}
\label{sizescaling}
B\simeq M^{3/4}.
\end{equation}
It is remarkable to note that all important demographic, socioeconomic,
and behavioral urban indicators such as consumption of energy and resources,
 production of artifacts, information, and waste,
 are, on average, scaling functions of city size that appear to be
very general to all cities, across urban systems
\cite{Bettencourt}.

The evolution of social and economic life in the
city increases with its population size: wages, income, growth
domestic product, bank deposits, as well as rates of invention,
measured by new patents and employment in creative sectors scale
super-linearly, over different years and nations with
statistically consistent exponents \cite{Florida}.

The probable reason for such a similarity is that highly complex,
self-sustaining structures, whether cells, organisms, or cities
constitute of an immense  numbers of units being organized in a
form of self-similar hierarchical branching networks, which grow
with the size of the organism \cite{Enquist}. A universal social
dynamic underlying the scaling phenomena observed in cities
implies that an increase in productive social opportunities, both
in number and quality,
 leads to quantifiable changes in individual
behavior of humans in a city integrating them into a complex dynamical
network \cite{Macionis}.

In the present paper, we consider the dual graph representation of urban texture.
After a brief introduction into city
space syntax (Sec.~\ref{sec:Graphrepresentations}) we
describe the typical space syntax measures (Sec.~\ref{sec:Analyticmeasures}) and
then investigate the statistics of
their values
for the five different compact urban patterns (Sec.~\ref{sec:complex_network}).
Universal statistical behavior of space syntax measures
 uncovers the universality of
the creation
mechanism responsible for the
 appearance of
nodes of high centrality which acts over
all cities independently
of their backgrounds.

\section{Graphs and space syntax of urban environments}
\label{sec:Graphrepresentations}
\noindent

Urban space is of rather large scale
 to be seen from a single viewpoint;
maps provide us with its representations by
means of abstract symbols facilitating our
 perceiving and understanding of a city.
 The middle scale and small scale
maps are usually based on Euclidean geometry
 providing spatial objects with precise coordinates
 along their edges and outlines.

There is a long tradition of research articulating urban
environment form using graph-theoretic
principles originating from the paper of Leonard Euler
(see \cite{GraphTheory}).
Graphs have long been regarded as the basic structures
for
representing forms where topological relations are firmly
 embedded within Euclidean
space.
The widespread use of graph theoretic analysis in geographic
 science had been reviewed in \cite{Haggett}
establishing it as central to spatial analysis of urban environments.
In \cite{Kansky}, the basic graph
theory methods had been applied  to the measurements of
 transportation networks.

Network analysis has long been a basic function of geographic
information systems
(GIS) for a variety of applications, in which computational
 modelling of an urban network is based on a graph view in
 which the intersections of linear features are regarded as
nodes, and connections between pairs of nodes are represented
as edges \cite{MillerShaw}.
Similarly, urban forms are usually represented as the patterns
of identifiable urban elements such as
locations or areas (forming nodes in a graph) whose relationships
to one another are often associated with linear
transport routes such as streets within cities \cite{Batty}.
Such planar graph representations define locations or points in
Euclidean plane as nodes or vertices $\{ i\}$, $i=1,\ldots, N$, and
 the edges linking  them together as $i\sim j$, in
 which $\{i,j\}=1,2,\ldots,N.$ The value of a link can
 either be binary, with the value $1$ as $i\sim j$, and
 $0$ otherwise, or be equal to actual physical distance
 between nodes,  $\mathrm{dist}(i,j)$, or to some weight $w_{ij}>0$ quantifying
a certain characteristic property of the link.
We shall call a planar graph representing the Euclidean space
embedding of an urban network as its {\it primary graph}.
 Once a spatial system has been identified and
represented by a graph in this way, it can be subjected to
the graph theoretic analysis.

A {\it spatial network} of a city is a network of the spatial
 elements of urban environments. They
are derived from maps of {\it open spaces} (streets, places, and roundabouts).
Open spaces may be broken down into components; most simply, these
might be street segments, which can be linked into a network via
their intersections and analyzed as a networks of {\it movement
choices}. The study of spatial configuration is instrumental in
predicting {\it human behavior}, for instance, pedestrian
movements in urban environments \cite{Hillier96}. A set of
theories and techniques for the analysis of spatial configurations
is called {\it space syntax} \cite{Jiang98}. Space syntax is
established on a quite sophisticated speculation that the
evolution of built form can be explained in analogy to the way
biological forms unravel \cite{SSyntax}. It has been developed as
a method for analyzing space in an urban environment capturing its
quality
 as being comprehendible and easily navigable \cite{Hillier96}.
Although,  in its initial form, space syntax was focused mainly on
patterns of pedestrian movement in cities, later the  various
space syntax measures of urban configuration had been found to be
correlated with the different aspects of social life,
\cite{Ratti2004}.

Decomposition of a space map into a complete set of
intersecting axial lines,  the fewest and
longest lines of sight that pass through every open space comprising any system,
produces an axial map or an overlapping convex map respectively.
Axial lines and convex spaces may be treated as the {\it spatial elements}
 (nodes of a morphological graph),
 while either the {\it junctions} of axial lines or the {\it overlaps} of
 convex  spaces may be considered as the edges linking  spatial elements
 into a single  graph unveiling the
topological relationships  between all open elements of the urban space.
In what follows,  we shall call this morphological representation of urban network
as a {\it dual graph}.

The transition to a dual graph is a topologically non-trivial
transformation of a planar primary graph into a non-planar one which
encapsulates the hierarchy and structure of the urban area and also corresponds
 to perception of space that people experience when
travelling along routes through the environment.

In Fig.~1, we have presented the glossary establishing a correspondence
 between several typical elements of urban environments and the certain subgraphs
 of dual graphs.
The dual transformation replaces the 1D open segments (streets) by the
 zero-dimensional nodes, Fig.~1(1).
\begin{figure}[ht]
\label{F1_11}
 \noindent
\begin{center}
\begin{tabular}{llrr}
 1. &\epsfig{file=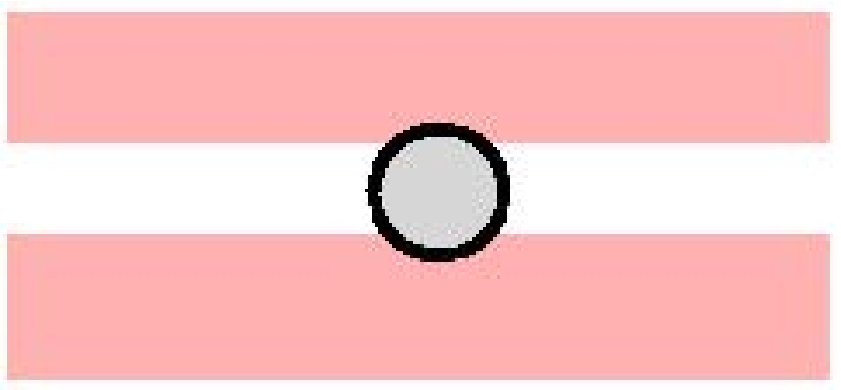, width=3.0cm, height =1.5cm}&2.&
 \epsfig{file=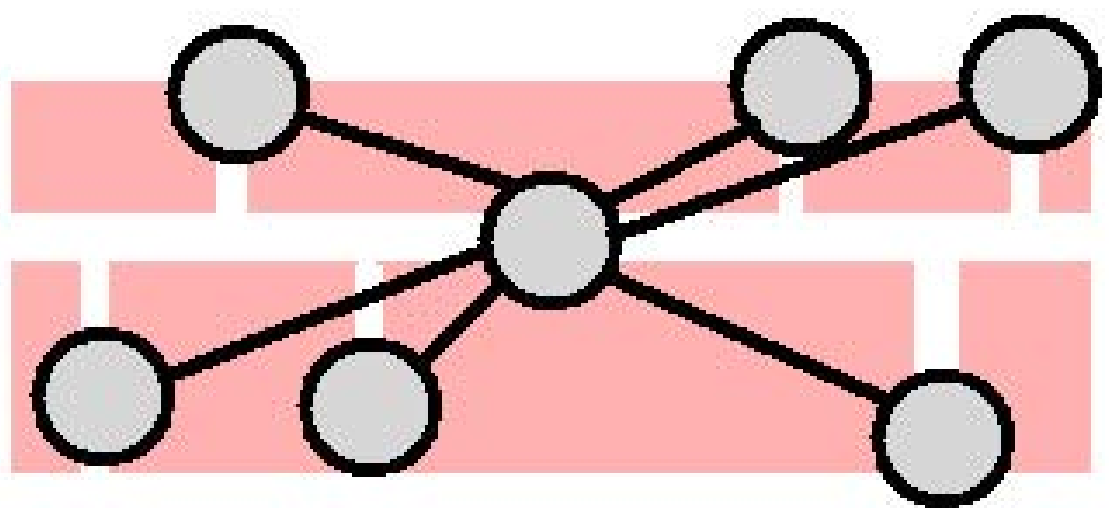, width=3.0cm, height =1.5cm} \\
 3. &\epsfig{file=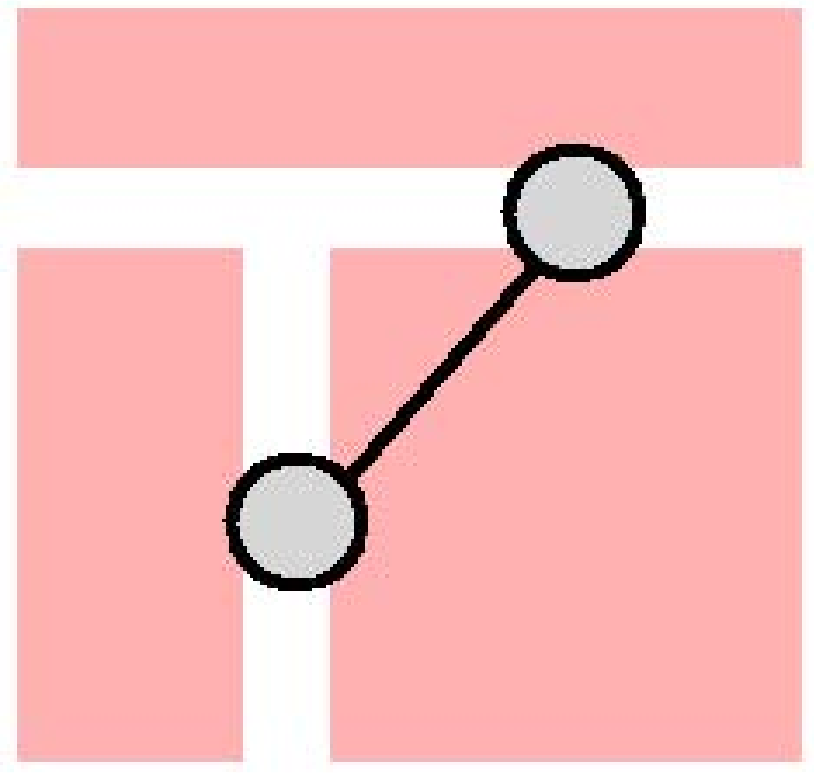,width=3.0cm, height =3.0cm}&4.&
 \epsfig{file=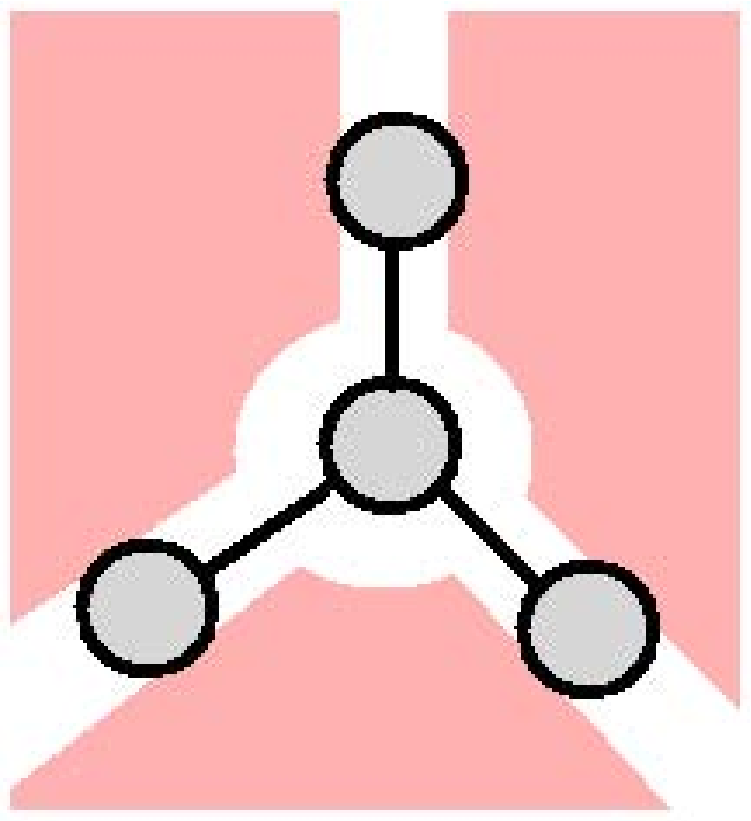, width=3.0cm, height =3.0cm} \\
 5. &\epsfig{file=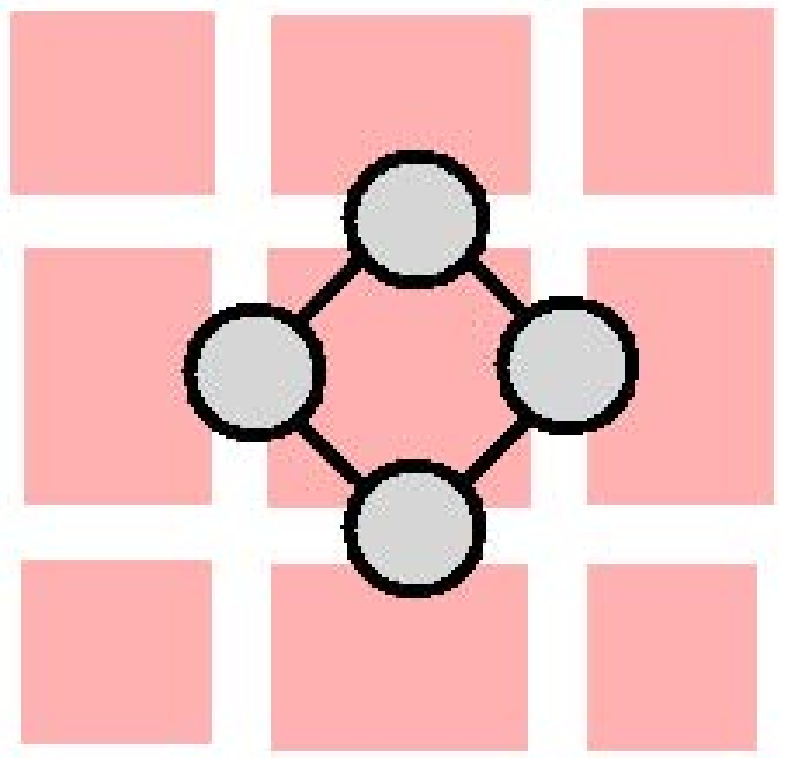,width=3.0cm, height =3.0cm}&6.&
 \epsfig{file=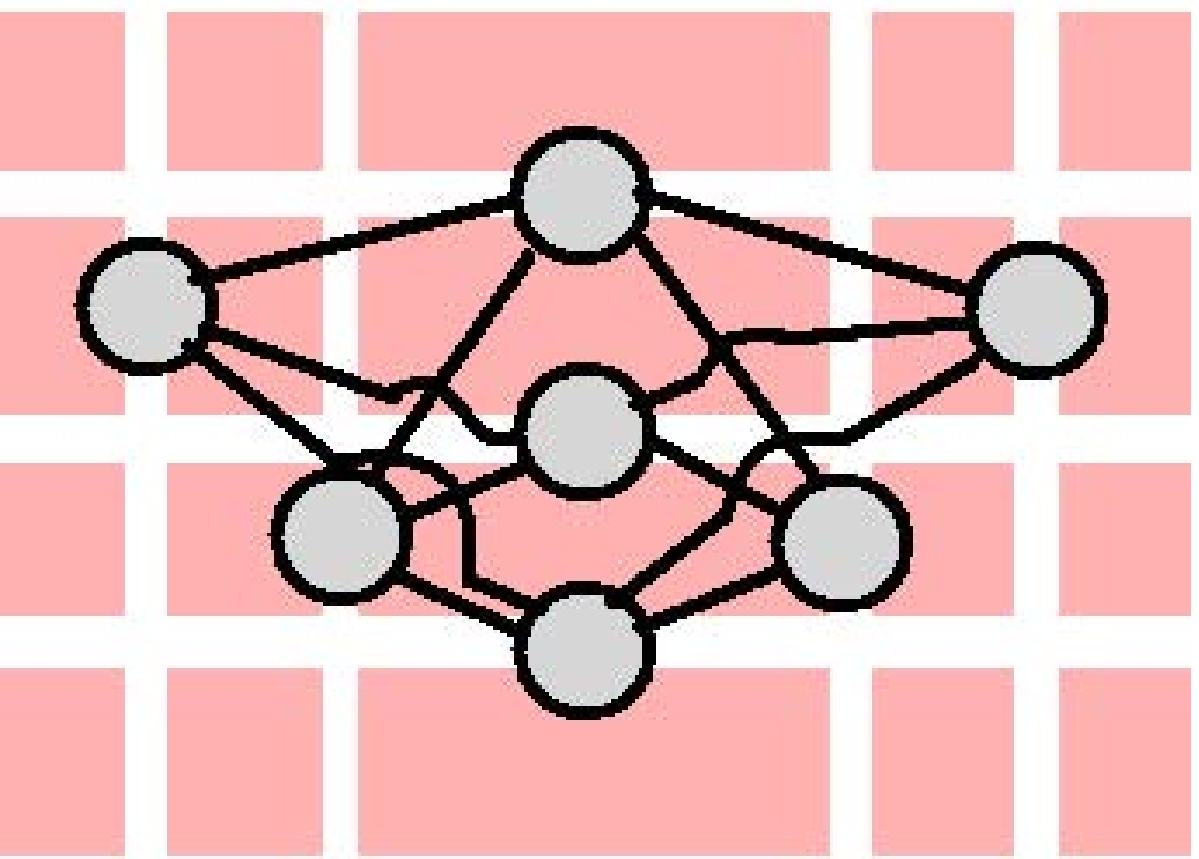, width=3.0cm, height =3.0cm} \\
\end{tabular}
\end{center}
\caption{The dual transformation glossary between the typical elements of
urban environments and the certain subgraphs of dual graphs.}
\end{figure}

The sprawl like developments consisting of a number of blind passes branching
 off a main route are changed to the star subgraphs having a hub and a number
 of client nodes, Fig.~1(2). Junctions and crossroads are replaced with
 edges connecting the corresponding nodes of the dual graph, Fig.1(3).
  Places and roundabouts are considered as the independent topological
  objects and acquire the individual IDs being nodes in the dual
  graph Fig.~1(4). Cycles are converted into  cycles of
  the same lengthes Fig.~1(5). A regular grid pattern
  is shown in Fig.~1(6). Its dual graph representation
  is called a {complete bipartite} graph, where the set of vertices
  can be divided into two disjoint subsets  such that no edge has
   both end-points in the same subset, and every line joining the
   two subsets is present, \cite{Krueger89}.
These sets can be naturally interpreted as those of the vertical
and horizontal edges in the primary graphs (streets and avenues).
In bipartite graphs, all closed paths are of even length, \cite{Skiena}.

 It is the dual
graph transformation which allows to separate the effects of order
and of structure while analyzing a transport network on the
morphological ground. It converts the repeating geometrical
elements expressing the order in the urban developments into the
{\it twins nodes}, the pairs of nodes such that any other is
adjacent either to them both or to neither of them. Examples
of twins nodes can be found in Fig.~1(2,4,5,6).

\section{Space syntax measures}
\label{sec:Analyticmeasures}
\noindent

A number of configurational {\it measures}
have been introduced in so far
in quantitative representations
of relationships between
spaces of urban areas and buildings.
Below we give a
brief introduction into
the measures commonly accepted in
 space syntax theory.

Although similar parameters
quantifying
connectivity and centrality  of nodes in a graph
have been independently invented and
extensively studied during
the last century in a varied range of disciplines
including
computer science, biology, economics, and sociology, the
syntactic measures are by no means just the new names for the well
known quantities.
In space syntax, the spaces are
understood as voids between
buildings restraining
traffic  that dramatically changes their meanings and the
interpretation of results.

Space {\it adjacency} is a basic rule to form axial maps:
two axial lines
 intersected are regarded as adjacency. Two spaces, $i$ and $j$, are
held to be {\it adjacent} in the dual graph $\mathfrak{G}$
when it is possible
to {\it move freely} from one
space to another, without passing through any
intervening.

The {\it adjacency matrix}  ${\bf A}_\mathfrak{G}$ of the dual graph
   $\mathfrak{G}$ is defined as follows:
\begin{equation}
\label{adjacencymatrix}
({\bf A}_\mathfrak{G})_{ij}=\left\{
\begin{array}{ll}
1,& i\sim j, \\
0,& \mathrm{otherwise}.
\end{array}
\right.
\end{equation}
Let us note that
rows and columns of ${\bf A}_\mathfrak{G}$
corresponding to
 the twins nodes are identical.
{\it Depth} is a topological distance
between nodes
in the dual graph $\mathfrak{G}$.
Two open spaces,  $i$  and $j$, are said to
 be at  depth $d_{ij}$ if
the
{\it least number} of syntactic steps
needed to reach one node from the other
is $d_{ij}$, \cite{glossary}.
The concept of depth can be extended to {\it total depth},
the sum of all
depths from a given origin,
\begin{equation}
\label{total_depth}
\mathfrak{D}_i=\sum_{j=1}^N d_{ij},
\end{equation}
in which $N$ is the total number of nodes in  $\mathfrak{G}$.
The average number of syntactic
steps from a given node $i$ to any other node
in the dual graph $\mathfrak{G}$
is called the
{\it mean depth},
\begin{equation}
\label{meandepth}
\ell_i=\frac {\mathfrak{D}_i}{N-1}.
\end{equation}
 The mean depth (\ref{meandepth}) is
 used for
quantifying the level of integration/segregation of the given
node,
\cite{Jiang98}.

{\it Connectivity}
is defined in space syntax theory as the number of nodes that
connect directly to a given node in the dual graph $\mathfrak{G}$, \cite{glossary}.
In graph theory, the space syntax connectivity\footnote{In
 graph theory, connectivity of a node is defined as the
{\it number of edges} connected to a vertex. Note that it
  is not
necessarily equal to the degree of node,  $\deg(i)$,  since there may be
more than one edge between any two vertices in the graph.}
 of a node is called the
 node  {\it degree}:
\begin{equation}
\label{connectivity}
\begin{array}{lcl}
\mathrm{Connectivity}(i) & =& \deg(i)\\
                         & = & \sum_{j=1}^N({\bf A}_\mathfrak{G})_{ij} .
\end{array}
\end{equation}
The {\it accessibility} of a space is considered in space syntax
as a key determinant of its spatial interaction and its analysis
is based on an implicit graph-theoretic view of the dual graph.

{\it Integration} of a node
is by definition expressed by a value that indicates the
degree to which a node is integrated or segregated from
a system as a whole ({\it global integration}), or from a partial
system consisting of nodes a few steps away ({\it local
integration}), \cite{JAG}.
It is measured by the {\it Real Relative Asymmetry} (RRA) \cite{Krueger89},
\begin{equation}
\label{integration2}
\mathrm{RRA}(i)\,=\, 2\,\frac{\ell_i -1}{D_N\,(N-2)},
\end{equation}
in which the normalization parameter
allowing to compare nodes belonging
to the dual graphs of different sizes
 is
\begin{equation}
\label{D_value}
D_N= 2\frac{N\left(\log_2\left(\frac{N+2}{3}\right)-1\right)+1}
{(N-1)(N-2)}.
\end{equation}
Another local measure used in  space syntax theory is the
{\it control value} (CV). It evaluates the degree
to which a space controls access to its immediate
neighbors taking
into account the number of alternative connections that each of
these neighbors has. The control value is
determined according to the following
formula, \cite{Jiang98}:
\begin{equation}
\label{controlvalue}
\begin{array}{lcl}
\mathrm{CV}(i) &=& \sum_{i\sim j}\frac 1{\deg(j)}\\
 &=& \sum_{j=1}^N\left({\bf A}_\mathfrak{G}{\bf D}^{-1}\right)_{ij}\\
\end{array}
\end{equation}
where the
diagonal matrix is
$
{\bf D}=\mathrm{diag}\left(\deg(1),\deg(2),\ldots,\deg(N)\right).
$

A dynamic global measure
of the {\it flow} through a space $i\in \mathfrak{G}$ commonly accepted
in  space syntax theory  is the
{\it global choice}, \cite{Hillier1987}. It captures how often a node may
be used in journeys from all spaces to all others
spaces in the city.
Vertices that occur on many shortest paths between other vertices have higher
 betweenness than those that do not.
Global choice can be estimated as the ratio between the number
 of shortest paths through the node $i$ and the total number of
  all shortest paths in $\mathfrak{G}$,
\begin{equation}
\label{globalchoice}
\mathrm{Choice}(i)=
\frac{\{\# \mathrm{ shortest {\ } paths {\ } through {\ }} i \}}
{\{\# \mathrm{ all {\ }shortest {\ } paths\}}}.
\end{equation}
A space $i$ has a
{\it strong choice} value when many of the shortest
paths, connecting all spaces to all spaces of a
system, passes through it.
 The Dijkstra's
classical  algorithm \cite{Djikstra}
which visits all nodes that are closer to the source
than the target before reaching the target
can be implemented in order to compute the value $\mathrm{Choice}(i)$.

The integration and the global choice index are the
centrality measures which capture the relative structural importance of a
node in a dual  graph.

\section{Space syntax as a complex network}
\label{sec:complex_network}
\noindent

The encoding of cities into non-planar dual graphs
reveals their complex
structure, \cite{Figueiredo2007}.

In order to illustrate the applications of complex network theory
methods to the dual graphs of urban environments, we have studied
 five different compact urban patterns.

 Two of them are
situated on islands: Manhattan (with an almost regular grid-like
city plan) and the network of Venice canals
(imprinting the joined
effect of natural, political, and economical factors acting on the
network during many centuries).
In the old city center
of Venice that stretches across 122 small islands in the marshy
 Venetian Lagoon along the Adriatic Sea in northeast Italy,
 the canals serve the function of roads.

We have also considered two organic cities
founded shortly after the Crusades and developed within the medieval
fortresses: Rothenburg ob der Tauber, the medieval Bavarian city
preserving its original structure from the 13$^\mathrm{th}$ century,
and the downtown of
Bielefeld  (Altstadt Bielefeld),
 an economic and cultural center of Eastern Westphalia.

To supplement the study of urban canal networks, we have
investigated that one in the city of Amsterdam. Although
it is not actually isolated from the national canal network, it is
binding to the delta of the Amstel river, forming a dense canal
web exhibiting a high degree of radial symmetry.

The scarcity of
physical space is among the most important factors determining the
structure of compact urban patterns. Some characteristics of studied dual city graphs are
given in Tab.1. There, $N$ is the number of open spaces (streets/canals and places)
 in the urban pattern
(the number of nodes in the dual graphs), $M$ is the number of junctions (the number of
edges in the dual graphs);
the graph {\it diameter},
$\mathrm{diam}(\mathfrak{G})$ is the {\it maximal} depth (i.e., the graph-theoretical distance)
between two vertices in a dual graph; the intelligibility parameter
estimates
navigability of the city, suitability for the passage through it.

In the framework of  complex network theory, the focus of study
is shifted away from the analysis of properties of individual
vertices to consideration of the {\it statistical} properties of
graphs, \cite{Newman2003}.

The
{\it degree distribution} has become an important
 concept in complex network theory
 describing the topology of complex networks.
It originates from
 the study of random graphs by Erd\"{o}s and R\'{e}nyi, \cite{ErdosRenyi}.
The degree distribution is
the probability that a node selected at random
among all nodes of the graph
 has exactly $k$ links,
\begin{equation}
\label{degdistr01}
P(k)=\Pr\left[i\in \mathfrak{G}|\deg(i)=k\right].
\end{equation}
The probability degree distributions for the
dual graph representations of the five compact urban patterns
mentioned in Tab.~1 has been studied by us in \cite{Volchenkov2007}.
 It is remarkable that the
observed profiles are broad
indicating that a street in a compact
city can cross the different number of other streets,
 in contrast while
in a regular grid.
At the same time,
the distributions usually
 have a clearly noticeable maximum
corresponding to the most probable
number of junctions an average street has in the city.
The long right tail of the distribution
which could decay even faster then a power law
is due to just a few "broadways," embankments,
and belt roads crossing many more streets than
an average street in the city, \cite{Volchenkov2007}.
It has been suggested in
\cite{Figueiredo2007} that irregular shapes
 and faster decays in
the tails of degree statistics
indicate that
the connectivity distributions  is
{\it scale-dependent}.
As a possible reason for the
  non-universal behavior is that
 in
the mapping and descriptive procedures,
inadequate choices to determine the
 boundary of the maps or flaws
in the aggregation process can damage
the representation of very
long lines.
Being scale-sensitive in general,
the degree statistics of dual city graphs
 can nevertheless be approximately
universal within a particular range
of scales.

\subsection{Integration  statistics of dual city graphs }
\label{subsec:Integration_statistics}
\noindent

The integration level of a node can be estimated by means of the
Real Relative Asymmetry index (\ref{integration2}).
It is obvious that if an urban network has a clear
hierarchical structure,
just a few of its nodes is well integrated becoming
 a part of the city core.
 Most of other routes are less integrated
growing like branches from the
main routes and penetrating inside
the quarters, but they do not
penetrate too much inside the "flesh" of the city.

\begin{figure}[ht]
 \noindent
\begin{center}
\epsfig{file=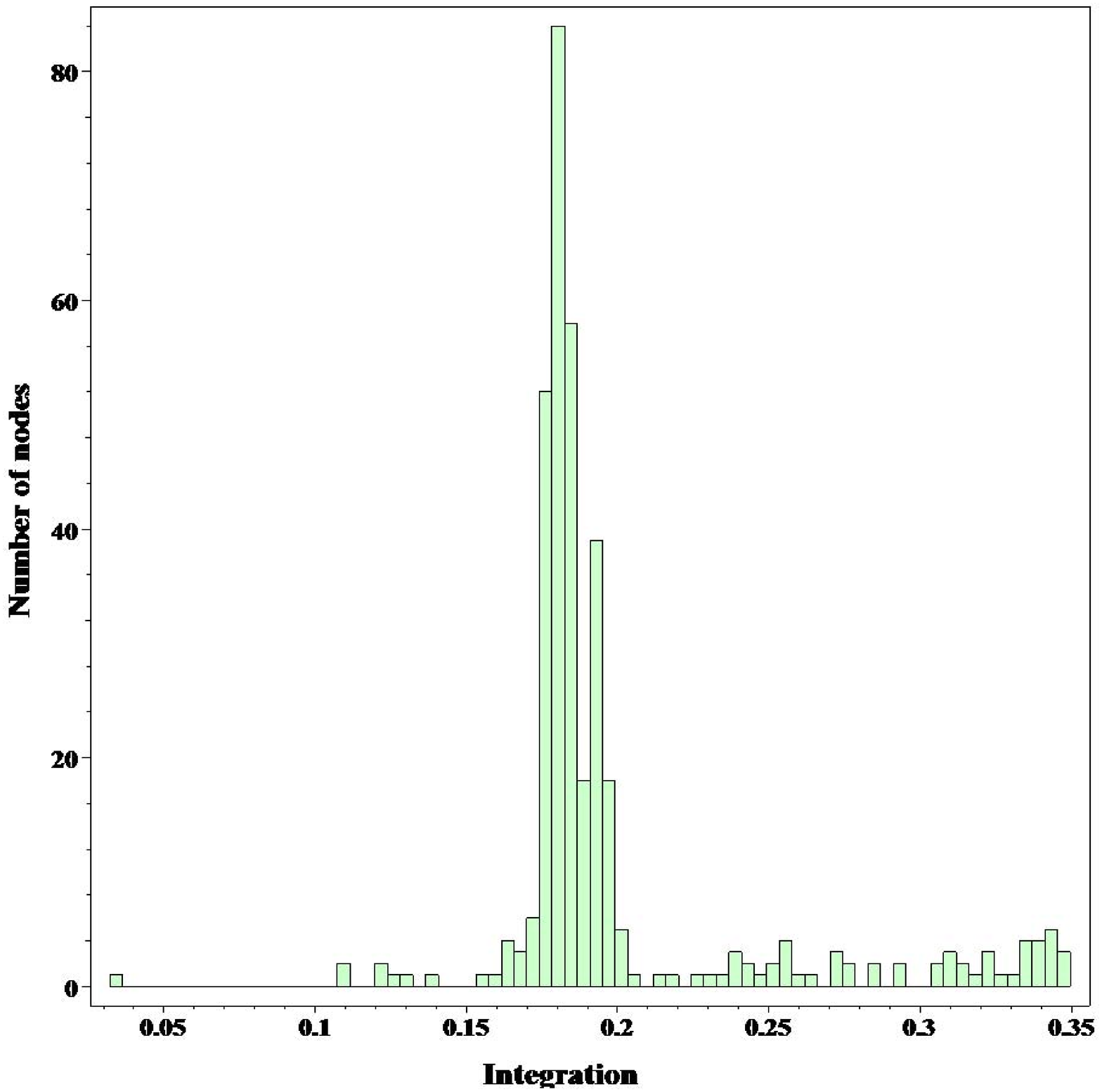, angle=0,width =7cm, height =6cm}
  \end{center}
\caption{\small The histogram shows the  distribution of nodes in
the dual graph representing the morphology of the street grid in
Manhattan over the Real Relative Asymmetry index
(\ref{integration2}) $\mathrm{RRA}(i)\in[0,1]$.}
 \label{Fig1_12}
\end{figure}

The distribution of nodes over the integration values can be
  represented in the form of a histogram (Fig.~\ref{Fig1_12}).
The height of each bar in the
histogram is equal to the number of
 nodes in the dual graph for which
the integration values  fall into the
 disjoint intervals (known as bins).
There is no {\it apriority} the best number of bins for the data
sample, and the histograms drawn
with respect to the different sizes of bins
 can reveal different features
of the data. We have used the Scott's method
 \cite{Scott} in order to calculate the bin width $h$,
\begin{equation}
\label{Scotts}
h=3.5\frac{\sigma}{N^{1/3}},
\end{equation}
where $N$ is the graph size, and $\sigma$ is the standard deviation
 of the RRA data, $\mathrm{RRA}(i)$, $i\in\mathfrak{G}$.

The integration
histogram
Fig.~\ref{Fig1_12} has a sharp peak located
at $\mathrm{RRA}(i)=0.17$. It indicates that the valuable
fraction
 of
streets in Manhattan belong to a class of
secondary  routes
characterized by the relatively low $\mathrm{RRA}$ values.
The sharp decay of the
histogram
  at the utmost
right cut represents a few nodes
identified as an integration {\it city core}.
A small number of streets
that is at the utmost left position in Fig.~\ref{Fig1_12}
represent the segregated streets.
It is worth
to mention that in contrast
to organic cities (which we discuss below),
 where the center of
integration city core
 (the {\it syntactic center} of the city) usually
matches the geographical city center,
 the essentially strongly
integrated streets
 in the city of hierarchical structure are
either the "broadways" crossing the city (the Broadway in
Manhattan) or the belt routes
encircling the city perimeter.

The
mismatching of
integration
profiles
can be used
for
the purpose
of comparison between the different
urban networks provided the correspondent
integration histograms are
normalized
with respect to the network sizes.
\begin{figure}[ht]
 \noindent
\begin{center}
\epsfig{file=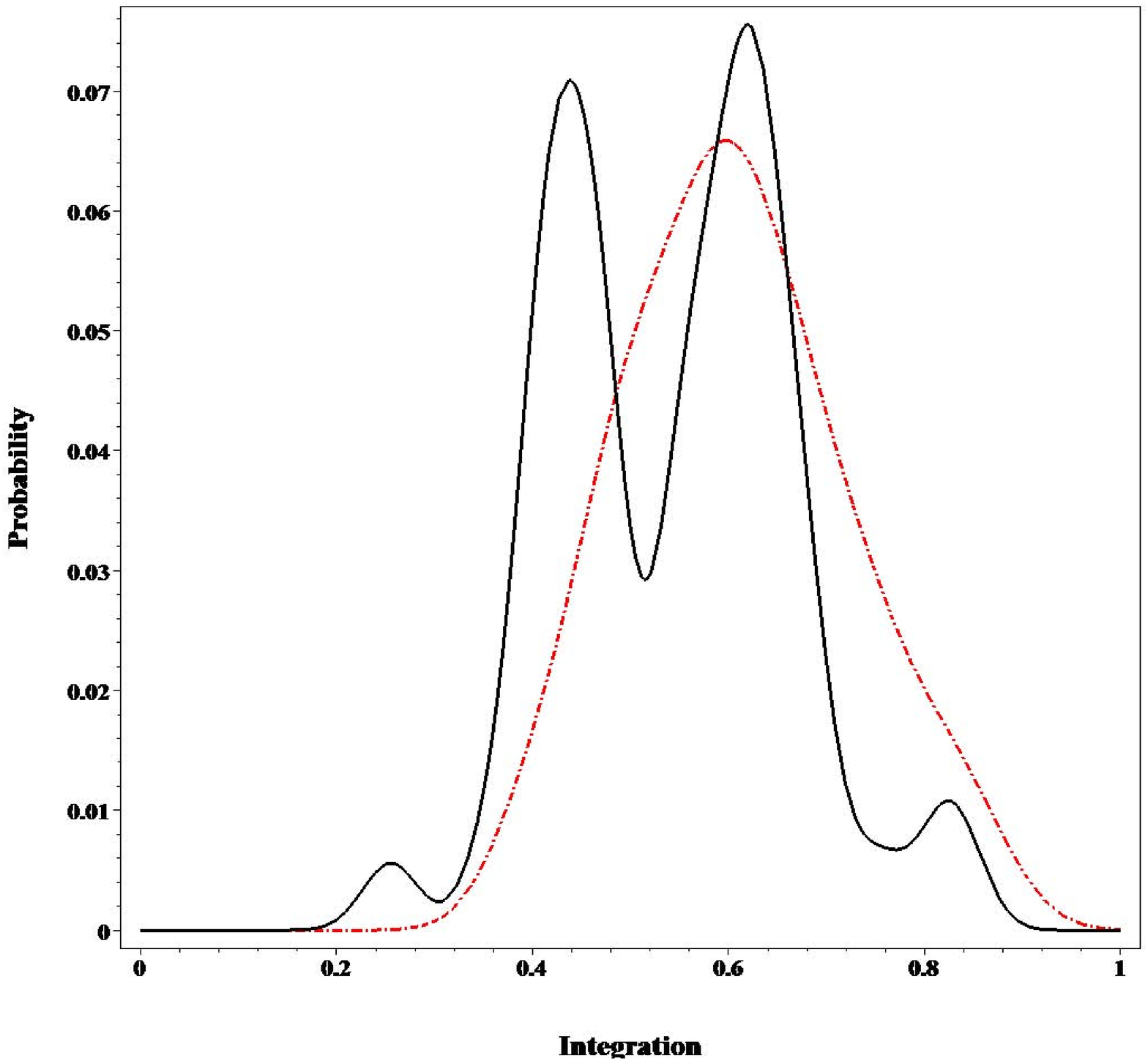, angle=0,width =8cm, height =6cm}
  \end{center}
\caption{\small The continuous approximations of the empirical
probability distributions $P(\mathfrak{i})$ calculated for two
German medieval cities: Rothenburg ob der Tauber,
 (the red dashed line) and the city of Bielefeld
(the black solid line). While the city of Rothenburg preserves its
original organic structure from the late $13^\mathrm{th}$ century,
the Bielefeld downtown is
composed of two different parts: the old section founded in the
$13^\mathrm{th}$  century and another one
subjected
to partial
urban  redevelopment in modern times.
 The structural difference between two
cities is clearly visible on their
 integration statistics profiles.}
\label{Fig1_13}
\end{figure}

Let us consider the probability distribution for the RRA
index $P(\mathfrak{i})$ such that a node $i$ selected at
random among all nodes of the dual graph $\mathfrak{G}$ has
the value $\mathrm{RRA}(i)=\mathfrak{i}$,
\begin{equation}
\label{probabintegr}
P(\mathfrak{i})=\Pr \left[i\in\mathfrak{G}|\mathrm{RRA}(i)=\mathfrak{i}\right].
\end{equation}
The continuous approximations for the empirical probability
distributions $P(\mathfrak{i})$ calculated for two German medieval
cities subjected to the organic development unveil their
structural difference.

The integration cores of the old organic cities usually form a
compact, dense sub-structure. The empirical probability
distributions for such cities are typically bell shaped (see the
red dashed line representing the continuous approximation of
$P(\mathfrak{i})$ for the city of Rothenburg). The major routes of
such a city are well integrated becoming a part of the city core
while the secondary routes branch from them penetrating inside the
quarters, but not too much inside of the city 'bulk'.

The empirical probability distribution for the city of Bielefeld
(see the black solid line in Fig.~\ref{Fig1_13})
 is of essential interest since its downtown is comprised of two
structurally different parts. The
ancient part of the city downtown keeps its
original organic structure
from
the late $13^\mathrm{th}$  century, while another part
was subjected to the partial redevelopment
twice, at the end of the $19^\mathrm{th}$ and in the
 $20^\mathrm{th}$ century.

The profiles of integration statistics
clearly display structural differences
between two urban patterns.
%
The empirical integration distribution
for the city of Bielefeld has
 two maxima that is an evidence in
favor of two different
  {\it most probable} integration
values.
It appears due to the fact that the
values of local
  integration
are  higher
for streets
belonging to one and the
  same city district,
while it is lower with respect to
those streets from alternative
city components.
%
The only city
    route ({\it Niederwall}) being a boundary between the ancient
    and redeveloped parts of the city conjugates them both.


Another, probably more intuitive measure
quantifying centrality of a node in a dual graph is the
global choice (\ref{globalchoice}).

The global choice value (\ref{globalchoice}) of a given node
counts the fraction of shortest paths between all origin/destination
pairs passing through it, $\mathrm{Choice}(i)\in [0,1]$.
Vertices that occur on many shortest paths between other
 vertices have higher global choice index than those that
 do not. The global choice analysis is preferred in network
  analysis to mean shortest-path length, as it gives higher
   values to more central vertices.
 It has recently been implemented in order to study and
 compare the differences between urban centers \cite{Crucitti}
 and for angular analysis within space syntax \cite{Shinichi}.
 The global choice is calculated by generating shortest paths
 between all nodes within the dual graph accordingly to the
  Dijkstra's algorithm \cite{Djikstra}.

We have investigated into the betweenness statistics of the dual
graphs of compact urban patterns. The {\it betweenness} is a
measure describing how shortest paths
 between the pairs of vertices are distributed over the network.
  The betweenness of vertex $i$ is nothing else but the probability
  that a randomly chosen shortest paths in the graph passes through $i$.
Most of vertices in
the dual graphs of the compact urban developments
are characterized by the relatively low values of the global choice
index (below, we call them as the 'bulk' nodes). At the same time just
a few nodes, the main city itineraries,
which
 play a key role in the overall city connectedness has the
  exceptionally high values of the global choice index.

\begin{figure}[ht]
 \noindent
\begin{center}
\epsfig{file=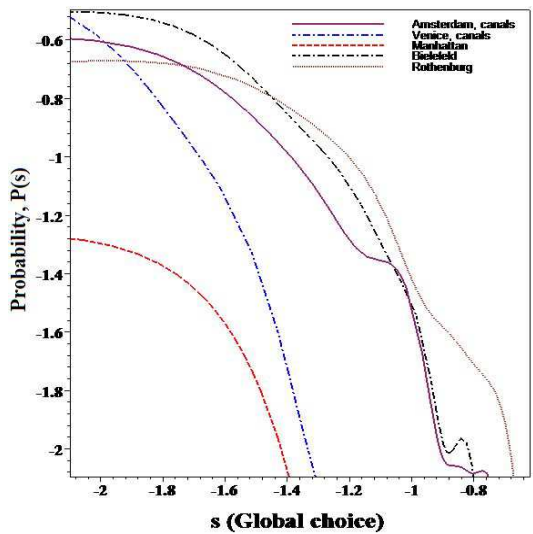, angle=0,width =7cm, height =7cm}
  \end{center}
\caption{\small The log-log plots of the continuous approximations
for the probability distributions of 'bulk' nodes over the value of
the global choice index $s$ calculated for
 five dual graphs of the compact urban patterns. }
 \label{Fig1_15a}
\end{figure}

Therefore a natural question arises on the distribution of the
'bulk' nodes in a dual graph over the global choice index
(betweenness). Let us consider the probability that a 'bulk' node
chosen randomly among all of them in the dual graph has the value
of the global choice index $s\in [0,1]$:
\begin{equation}
\label{globalprobab}
P(s)=\Pr\left[i\in \mathfrak{G}|\mathrm{Choice}(i)=s\right].
\end{equation}
In Fig.~\ref{Fig1_15a}, we have presented the log-log plot for the
continuous approximations for the  probability distributions of
the global choice index  for five dual graphs of the compact urban
patterns.
The obvious discrepancy in the probability distribution profiles
$P(s)$ (\ref{globalprobab}) indicate the multiple structural
differences between the dual graphs of the  studied urban
patterns. The probability that an arbitrary bulk node would have a
very small global choice index is high for the organic cities
(Bielefeld, Rothenburg, the canal networks of Venice and
Amsterdam), while it is considerably low for the city planned in a
regular grid (Manhattan).

The probability that a node has a strong global choice
 index decays very fast with the value  of $s$ for the
 modern street grid in Manhattan and for the eternal Venetian canals.
In both urban developments,
the strongest choice is
represented by just a few open spaces (Broadway, Franklin D
 Roosevelt Dr., West Side Hwy, and Henry Hudson Pkwy, in Manhattan;
 the Venetian Lagoon,
the Giudecca Canal, and the Grand Canal, in Venice)
while all other routes  contributing  to the statistics shown in Fig.~\ref{Fig1_15a}
play merely the subsidiary role canalizing the local flows towards the main itineraries.
In contrast to them, the probability distributions for the global choice index
over the open spaces in
 the compact cities of organic development decay much slower, with the
contributions from the main itineraries of the highest global choice indices
being a part of the continuous distributions.

\subsection{Control value  statistics of dual city graphs }
\label{subsec:CV_statistics}
\noindent

The $\mathrm{CV}(i)$-parameter
quantifies the degree of choice the node $i\in\mathfrak{G}$
represents for other
nodes directly connected to it.

\begin{figure}[ht]
 \noindent
\begin{center}
\epsfig{file=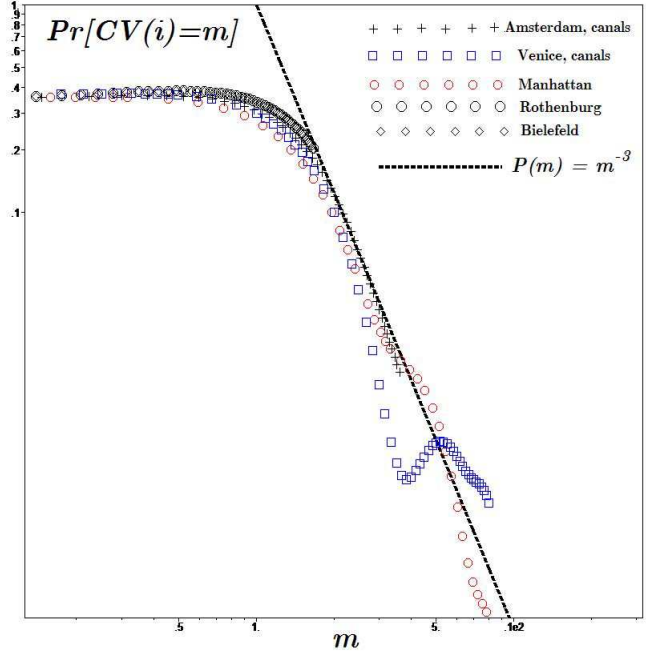, angle=0,width =9cm, height =9cm}
  \end{center}
\caption{\small The log-log plot of the probability distribution
that a node randomly selected among all nodes of the dual graph
 $\mathfrak{G}$ will be populated with precisely $m$ random walkers
  in one step starting from the uniform distribution (one random
   walker at each node). The dashed line indicates the cubic hyperbola
    decay, $P(m)=m^{-3}$.}
\label{Fig1_15}
\end{figure}

Provided random walks,
in which a  walker
moves in one step to
another node randomly chosen
 among all its nearest neighbors
 are
defined on the graph
 $\mathfrak{G}$, the
parameter $\mathrm{CV}(i)$
acquires a probabilistic interpretation.
Namely, it specifies
 the expected
number of
  walkers which is found in $i\in\mathfrak{G}$
after one step
if the random walks starts
 from a uniform configuration, in which
all nodes in the graph
have been  uniformly
populated
by precisely one walker.

Then, a graph $\mathfrak{G}$
can be characterized
by the probability
\begin{equation}
\label{CVdistr01}
P(m)=\Pr\left[i\in \mathfrak{G}|\mathrm{CV}(i)=m\right]
\end{equation}
of that the control value of a node chosen
uniformly at random among all nodes of the graph
$\mathfrak{G}$ equals to $m>0$.

The log-log plot of (\ref{CVdistr01}) is shown in  Fig.~\ref{Fig1_15}.
It is important to mention that
the profile of the probability decay
 exhibits the approximate scaling
well fitted by the cubic hyperbola,
  $P(m)\simeq m^{-3}$,
universally for all five compact urban
 patterns we have studied.

Universal statistical behavior of the
 control values for the nodes
representing a relatively
strong choice
for their nearest neighbors,
\begin{equation}
\label{CVstat}
\Pr\left[i\in \mathfrak{G}|\mathrm{CV}(i)=m\right] \simeq \frac 1{m^3},
\end{equation}
 uncovers the universality of
the creation
mechanism responsible for the
 appearance of
the "strong choice"
nodes which acts over
all cities independently
of their backgrounds.
 It
is a common suggestion in  space syntax theory
that  open spaces of strong choice
are
responsible for the {\it public space processes}
driven largely by the
universal social activities like trades
and exchange
which are common
across different cultures
and historical epochs
and  give cities
a similar global
structure of the "deformed wheel" \cite{SSBeijing}.

It has been shown long time ago by H. Simon \cite{Simon}
that the
power law distributions always arise when the {\it "the rich get richer"}
principle works, i.e. when a quantity increases with its amount already present.
In sociology this principle is known as the {\it Matthew effect} \cite{Matthew}
(this reference appears in \cite{Newman2003}) following the well-known biblical
edict.

\subsection{Intelligibility and Navigation}
\label{subsec:Intelligibility}
\noindent

The kinds of human behavior
that appear to be particularly
structured by open spaces are
pedestrian movement and navigation.
It appears that
the variance between
volumes of pedestrian
movement
at different
places in a city
 can be  predicted reasonably accurately from investigations
of spatial
configurations alone,
  \cite{Hillier1987}.
In space syntax,
 {\it correlations}
between local property
of a space (connectivity)
and global
configurational variables
(integration)
 constitute
a measure of the {\it intelligibility},
the global parameter
quantifying the part-whole
 relationship within the spatial
configuration.
Intelligibility
describes {\it how far} the
depth of a space from the street
 layout as a whole can be inferred
from the number of its direct connections \cite{glossary} that is
most important to way-finding  and perception of environments
\cite{Hillier92,Dalton_Intell}. More integrated areas were also
found to be more "legible" by the residents who perceived their
"neighborhood" to be of a greater size, \cite{Kim99}.

In the traditional space syntax approach, the
strong area definition and good intelligibility are
identified in an {\it intelligibility scattergram} and
then by means of the {\it Visibility Graph Analysis}
 (VGA) \cite{HillierEconomies}.
In statistics, a scatter plot is a useful summary of a set of two
variables, usually drawn before working out a linear correlation
coefficient or fitting a regression line. Each node of the dual
graph contributes one point to the scatter plot. The resulting
pattern indicates the type and strength of the relationship
between
 the two variables,
and aids the interpretation
 of the correlation coefficient \cite{Serfling}.

The measure of spatial integration for
each node $i\in\mathfrak{G}$ is usually
taken the
mean depth $\ell_i$,
 however
the  analysis varies for
specific case studies, and
the precise measures of
the graph are chosen
 to best correlate.

\begin{figure}[ht]
 \noindent
\begin{center}
\epsfig{file=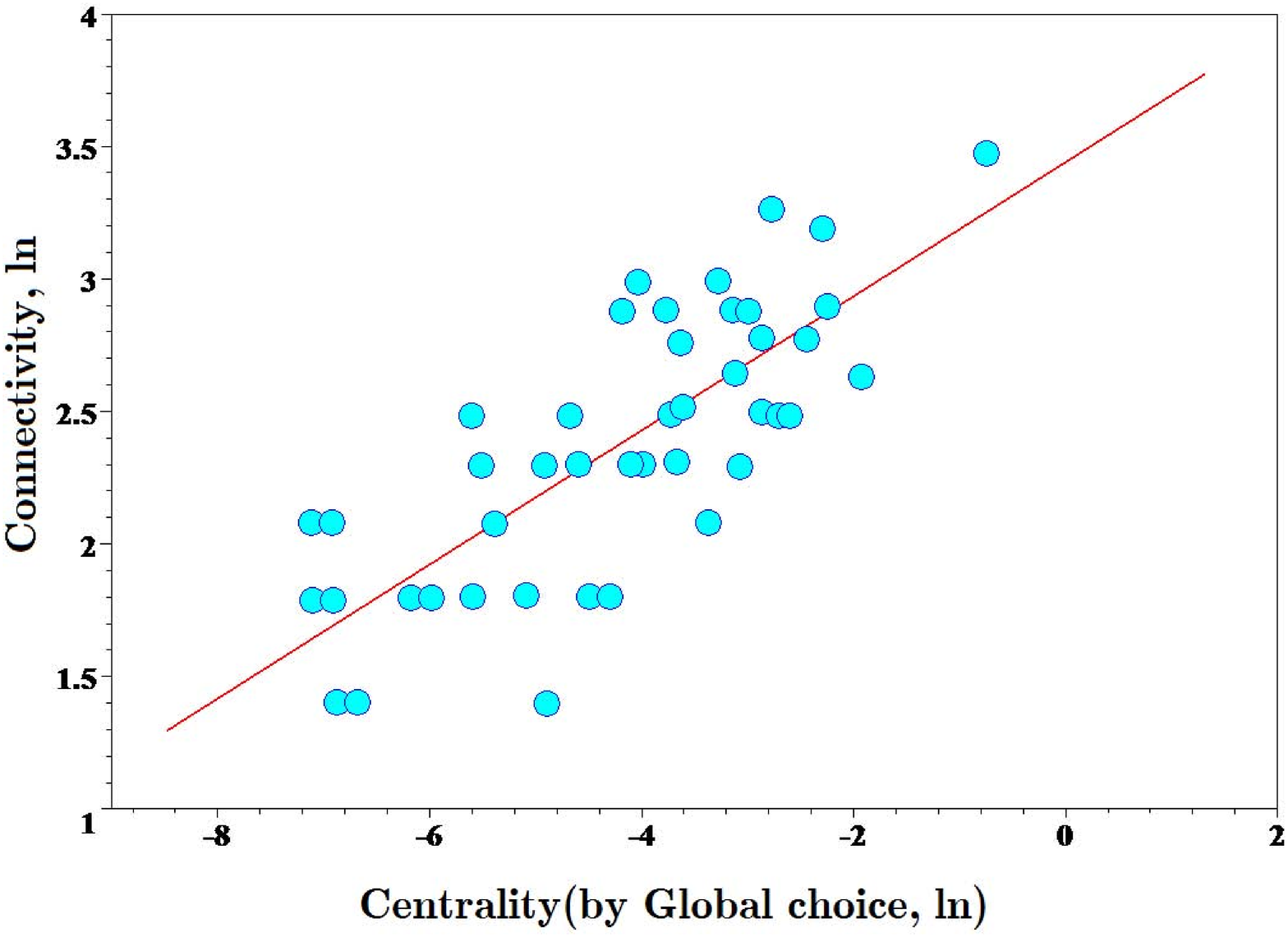, angle=0,width =8cm, height =6cm}
  \end{center}
\caption{\small The intelligibility scatter plot for the Bielefeld
downtown. The slope of the regression line fitting the data by the
method of least squares equals to $0.253$.} \label{Fig1_16}
\end{figure}

We prefer to use the global choice parameter (\ref{globalchoice})
as a measure of spatial integration. The scatter plot for the
downtown of Bielefeld (in the log-log scale) which shows the
relationship between connectivity and global choice (centrality)
 is
sketched on Fig.~\ref{Fig1_16}.
The pattern of dots (representing the certain open spaces
in the downtown of Bielefeld) slopes from lower left to upper right
that suggests a positive correlation between the connectivity and
centrality variables being studied.
A line of best fit
 computed using the method of linear regression
exhibits the slope $0.253$.
Let us note that the value of Pearson's
coefficient of linear correlations
between the  data samples
of connectivity and global choice
for Bielefeld equals to $0.681$.

\begin{figure}[ht]
 \noindent
\begin{center}
\epsfig{file=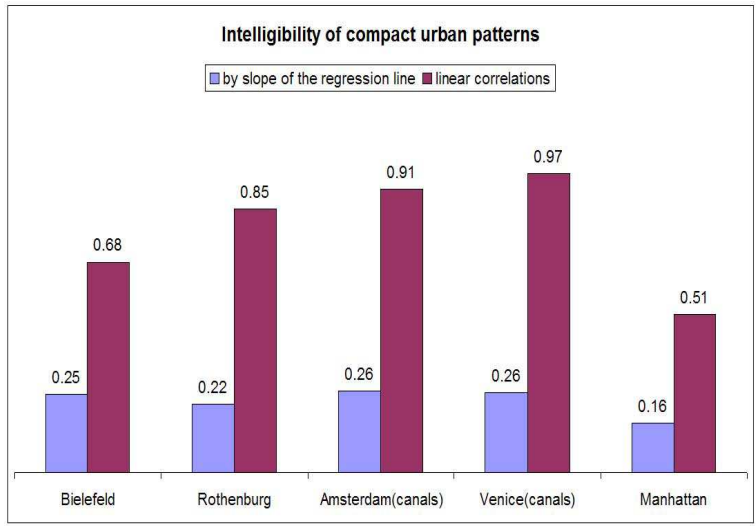, angle=0,width =10cm, height =7cm}
  \end{center}
\caption{\small The comparative diagram of intelligibility indices
calculated for the five compact urban patterns by different
methods.
 }
\label{Fig1_17}
\end{figure}

The correlations between local and global
properties within the spatial configurations
of urban networks
(intelligibility)
can be quantified by means of different methods.
The level of correlations can be reckoned by the
slope of the regression line fitting the data of the scatter plot
drawn in the logarithmic scale. Eventually, we can directly
compute  Pearson's
coefficient of linear correlations \cite{Pearson}
 between the uniformly ordered
connectivity and integration values. In order to show the
compatibility of  these methods, we collect the results of all
intelligibility estimations for the five compact urban patterns
that we studied in one diagram (see Fig.~\ref{Fig1_17}).
It is clear from the diagram (\ref{Fig1_17}) that
being an important characteristic related to
a perception of place and navigation within that,
intelligibility can be used in a purpose of
 comparison between
urban networks. The obvious advantage of
 intelligibility is that
it does not depend upon the network size.

The
intelligibility indices estimated by means
of Pearson's correlation
coefficient   have been given
 in Tab.~1.

\section{Conclusion}
\label{sec:Discussion}
\noindent

We have investigated the complex networks
of city space syntax observed in the several
compact urban patterns.
We have studied the local and various global measures
characterizing a node in a dual city graph as well as
the positive relationship
 between local and global properties known as intelligibility.

If cities were perfect grids in which all lines
 have the same length and number of
junctions, they would be described by
regular graphs exhibiting a
 high level of similarity no matter which part
of urban texture
 is examined. This would create a highly
accessible system that provides multiple routes between any pair of
 locations. It was believed that pure grid systems are easy to
  navigate due to this high
accessibility and to the existence of multiple paths between any
pair of locations \cite{RosvallPRE,RosvallPRL}. Although, the
urban grid minimizes descriptions as long as possible in the ideal
grid all routes are equally probable;
 morphology of the perfect grid does not differentiate main
spaces and movement tend to be dispersed everywhere \cite{Figueiredo2007}.
 Alternatively, if cities were purely hierarchical systems (like trees),
they would clearly have a main space (a hub, a single route between many
pairs of locations) that connects all branches and controls movement
between them.
 This would create a highly segregated, sprawl like system that would
 cause a tough social consequence, \cite{Figueiredo2007}.

However, real cities are neither trees nor perfect grids, but a
combination of these structures that emerge from the social and
constructive processes \cite{Hillier96}. They  maintain enough
differentiation to establish a clear hierarchy \cite{Hanson89}
resulting from a process of negotiation between the public
processes (like trade and exchanges) and the residential process
preserving their traditional structure. The emergent urban network
is usually of a very complex structure which is therefore
naturally subjected to the {\it complex network theory} analysis.

\section{Acknowledgment}
\label{Acknowledgment}
\noindent

The work has been supported by the Volkswagen Stiftung (Germany)
in the framework of the project: "Network formation rules, random
set graphs and generalized epidemic processes" (Contract no Az.:
I/82 418). The authors acknowledge the multiple fruitful
discussions with the participants of the workshop {\it Madeira
Math Encounters XXXIII}, August 2007, CCM - CENTRO DE CI\^{E}NCIAS
MATEM\'{A}TICAS, Funchal, Madeira (Portugal).

\newpage
\begin{center}
{\bf \small Table 1: Some features of studied dual city graphs}

\vspace{0.3cm}

\begin{tabular}{c|c|c|c|c}
   \hline \hline
 Urban pattern   & $N$ & $M$ & $\mathrm{diam}(\mathfrak{G})$ & Intelligibility
    \\ \hline\hline
 Rothenburg ob d.T. & 50 & 115 &  5& 0.85
 \\
Bielefeld (downtown)& 50 & 142 &  6& 0.68
 \\ 
 Amsterdam (canals) & 57 & 200 & 7& 0.91
\\
 Venice (canals) & 96 & 196 &  5& 0.97
 \\ 
  Manhattan & 355 & 3543 &  5& 0.51
 \\
  \hline \hline
\end{tabular}
\end{center}


\begin{thebibliography}{000}

\bibitem{Crane}
P. Crane, A. Kinzig,  {\it Science} {\bf 308}, 1225 (2005).
\bibitem{Miller}
J.G. Miller, {\it Living Systems}, McGraw-Hill, New York (1978).

\bibitem{Bettencourt}
L.M.A. Bettencourt, J. Lobo, D. Helbing, C. K\"{u}hnert, and G.B. West,
"Growth, innovation, scaling, and the pace of life in cities",
{\it PNAS} published online Apr 16, 2007; doi:10.1073/pnas.0610172104.

\bibitem{Florida}
R. Florida, {\it Cities and the Creative Class}, Routledge, New York, (2004).

\bibitem{Enquist}
B.J. Enquist, J.H. Brown, G.B. West, {\it Nature}  {\bf 395}, 163-166 (1998).

\bibitem{Macionis}
 J.J. Macionis,  V.N. Parillo,  {\it Cities and Urban Life}, Pearson Education,
Upper Saddle River, NJ (1998).

\bibitem{GraphTheory}
 N. Biggs,  E. Lloyd, and R. Wilson, {\it  Graph Theory, 1736-1936}.
 Oxford University Press (1986).

\bibitem{Haggett}
P. Haggett, R. Chorley (eds.), {\it Socio-Economic Models
in Geography}, London, Methuen (1967);

 P. Haggett, R. Chorley, {\it Network Analysis in Geography},
  Edward Arnold, London (1969).

\bibitem{Kansky}
 K.J. Kansky, {\it Structure of Transportation Networks: Relationships
Between Network Geometry and Regional Characteristics}, Research Paper {\bf 84},
Department of Geography, University of Chicago , Chicago, IL (1963).


\bibitem{MillerShaw}
H.J. Miller, S.L. Shaw, {\it Geographic Information Systems for
Transportation: Principles and Applications}, Oxford Univ. Press, Oxford (2001).

\bibitem{Batty}
M. Batty, {\it A New Theory of Space Syntax},
UCL Centre For Advanced Spatial Analysis Publications, CASA Working Paper
{\bf 75} (2004).


\bibitem{SSyntax}
 B. Hillier, A. Leaman, P. Stansall,  M.  Bedford,
{\it Environment and Planning B} {\bf 3}, 147-185 (1976).


\bibitem{Hillier96}
 B. Hillier, {\it Space is the machine. A configurational theory of
 architecture},  Cambridge University Press (1996).
\bibitem{Jiang98}
 B. Jiang, "A space syntax approach to spatial cognition in urban environments",
  Position paper for NSF-funded research workshop {\it Cognitive Models of Dynamic
   Phenomena and Their Representations}, October 29 - 31, 1998, University of
   Pittsburgh, Pittsburgh, PA (1998).

\bibitem{Ratti2004}
C. Ratti, {\it Environment and Planning B: Planning and Design},
 vol. {\bf 31}, pp. 487 - 499 (2004).

\bibitem{Krueger89}
 M.J.T. Kruger, {\it On node and axial grid maps: distance
 measures and related topics}. Other. Bartlett School of
 Architecture and Planning, UCL, London, UK (1989).


\bibitem{Skiena}
S. Skiena, {\it Implementing Discrete Mathematics: Combinatorics
 and Graph Theory with Mathematica}, Reading, MA: Addison-Wesley (1990).

\bibitem{glossary}
Bj\"{o}rn Klarqvist, "A Space Syntax Glossary", {\it Nordisk
Arkitekturforskning} {\bf 2} (1993).

\bibitem{JAG}
B. Jiang, Ch. Claramunt, and B. Klarqvist,
{\it JAG}, Vol. {\bf 2} (3/4), 161-171 (2000).


\bibitem{Hillier1987}
 B. Hillier, R. Burdett, J. Peponis,  A. Penn, {\it Architecture
  and Comportement / Architecture and Behaviour} {\bf 3}(3), 233-250 (1987).

\bibitem{Djikstra}
T. H. Cormen, C. E. Leiserson, R. L. Rivest, and C. Stein,
{\it Introduction to Algorithms}, Second Edition,
 Section 24.3: Dijkstra's algorithm, pp.595–601,  MIT Press and McGraw-Hill, (2001).


\bibitem{Figueiredo2007}
L. Figueiredo, L. Amorim, {\it Decoding the urban grid: or why
cities are neither trees nor perfect grids}, 6th International Space Syntax
Symposium, 12-15 Jun 2007, Istanbul, Turkey.

\bibitem{Newman2003}
 M.E.J. Newman, {\it SIAM Review} {\bf  45}, 167-256 (2003).
\bibitem{ErdosRenyi}
 P. Erdos, A. Renyi,  {\it Publ. Math.} (Debrecen) {\bf 6}, 290 (1959).
\bibitem{Volchenkov2007}
D. Volchenkov, Ph. Blanchard, {\it Phys. Rev. E}, {\bf 75}, 026104 (2007).

\bibitem{Crucitti}
P. Crucitti, V. Latora, and S. Porta, {\it Chaos} {\bf 16}, 015113 (2006).
\bibitem{Shinichi}
 B. Hillier, I. Shinichi, {\it Network effects and psychological effects:
  A theory of urban movement}, Proc. of the 5th International Symposium
  on Space Syntax Vol. {\bf 1}, TU Delft,Delft, Netherlands, pp 553–564 (2005).

\bibitem{Scott}
D.W. Scott, {\it Biometrika} {\bf 66} (3), 605–610 (1979).
\bibitem{SSBeijing}
B. Hillier,  "The art of place and the science of space", {\it World Architecture}
{\bf 11}/2005 (185), Beijing, Special Issue on Space Syntax pp. 24-34
 (in Chinese), pp. 96-102 (in English) (2005).

\bibitem{Simon}
 H. A. Simon, {\it Biometrika} {\bf 42}, 425–440 (1955).
\bibitem{Matthew}
R. K. Merton, {\it Science} {\bf 159}, 56–63 (1968).

\bibitem{Hillier92}
 B. Hillier, A.  Penn, N. Dalton,
"Milton Keynes: Looking back to London",
{\it The Architects
Journal}, 15th April 1992, London.
\bibitem{Dalton_Intell}
R. Dalton,
"Is Spatial Intelligibility
Critical to the Design of Large-scale Virtual Environments?",
{\it Int. J.  Design Computing} {\bf 4} (2002).



\bibitem{Kim99}
Y.O. Kim, {\it Spatial Configuration,
Spatial Cognition and Spatial Behaviour: the role of architectural
intelligibility in shaping spatial experience}, PhD Thesis, University of London (1999).
\bibitem{HillierEconomies}
B. Hillier, {\it Urban Design International}, {\bf 1}(1), 49-60 (1996).
\bibitem{Serfling}
R.J. Serfling. {\it Approximation Theorems of Mathematical Statistics}, John Wiley \& Sons
 (1980).
\bibitem{Pearson}
 J. Cohen,  P. Cohen,  S.G. West, L.S. Aiken,
  {\it Applied multiple regression/correlation analysis for the behavioral sciences}.
  (3rd ed.) Hillsdale, NJ: Lawrence Erlbaum Associates. (2003).

\bibitem{RosvallPRL}
M. Rosvall, A. Trusina, P. Minnhagen, K. Sneppen,
{\it Phys. Rev. Lett.} {\bf 94}, 028701 (2005).

\bibitem{RosvallPRE}
M. Rosvall, P. Minnhagen, K. Sneppen,
{\it Phys. Rev. E} {\bf 71}, 066111 (2005).


\bibitem{Hanson89}
J. Hanson, {\it Ekistics} {\bf 56}, pp. 334-335  (1989).


\end{thebibliography}
\end{document}